\documentclass{article}
\usepackage{xcolor}
\usepackage{amsmath} 
\usepackage{graphicx}
\usepackage[ruled,vlined]{algorithm2e}

\usepackage{array}
\newcolumntype{L}[1]{>{\raggedright\let\newline\\\arraybackslash\hspace{0pt}}m{#1}}
\newcolumntype{C}[1]{>{\centering\let\newline\\\arraybackslash\hspace{0pt}}m{#1}}
\newcolumntype{R}[1]{>{\raggedleft\let\newline\\\arraybackslash\hspace{0pt}}m{#1}}

\usepackage{colortbl}        

\usepackage{tabularx}
\usepackage{booktabs} 
\usepackage{ragged2e} 

\usepackage{hyperref}

\begin{document}

\title{Optimizing Federated Learning for Scalable Power-demand Forecasting in Microgrids\thanks{~Pre-print of paper to appear as part of IEEE International Conference on e-Science (eScience), 2025, Chicago. Paper was shortlisted as a Best Paper Finalist.}}

\author{%
Roopkatha Banerjee\textsuperscript{1}, Sampath Koti\textsuperscript{1}, Gyanendra Singh\textsuperscript{2},\\
Anirban Chakraborty\textsuperscript{1}, Gurunath Gurrala\textsuperscript{2},\\
Bhushan Jagyasi\textsuperscript{3}, Yogesh Simmhan\textsuperscript{1} \\[6pt]
\begin{minipage}{\linewidth}\centering\small
\textsuperscript{1}Department of Computational and Data Sciences,\\
Indian Institute of Science (IISc),\\
Bangalore, India\\
\textsuperscript{2}Department of Electrical Engineering,\\
Indian Institute of Science (IISc),\\
Bangalore, India\\
\textsuperscript{3}Accenture, India\\
\{roopkathab, simmhan\}@iisc.ac.in
\end{minipage}
}
\date{}
\maketitle
\begin{abstract}
Real-time monitoring of power consumption in cities and micro-grids through the Internet of Things (IoT) can help forecast future demand and optimize grid operations. 
But moving all consumer-level usage data to the cloud for predictions and analysis at fine time scales can expose activity patterns. Federated Learning~(FL) is a privacy-sensitive collaborative DNN training approach that retains data on edge devices, trains the models on private data locally, and aggregates the local models in the cloud. 
But key challenges exist: (i) clients can have non-independently identically distributed~(non-IID) data, and (ii) the learning should be computationally cheap while scaling to 1000s of (unseen) clients. In this paper, we develop and evaluate several optimizations to FL training across edge and cloud for time-series demand forecasting in micro-grids and city-scale utilities using DNNs to achieve a high prediction accuracy while minimizing the training cost. We showcase the benefit of using exponentially weighted loss while training and show that it further improves the prediction of the final model. Finally, we evaluate these strategies by validating over 1000s of clients for three states in the US from the OpenEIA corpus, and performing FL both in a pseudo distributed setting and a Pi edge cluster. The results highlight the benefits of the proposed methods over baselines like ARIMA and DNNs trained for individual consumers, which are not scalable.
\end{abstract}

\section{Introduction}
\label{sec:Introduction}
\textit{Smart power grids} have become popular to help manage the complexity and scale of electricity distribution from producers to consumers due to their real-time monitoring along with two-way communication for autonomous control~\cite{sehrawat2019smart}. Power usage data from consumers, sampled every $1$--$60~$mins from smart meters, is collected in public or private (utility) clouds to learn and predict energy usage patterns and cumulative demand, and respond to these to achieve efficient and reliable electrical power delivery for daily operations~\cite{da2023resource,10.1109/MCSE.2013.39}. Within a \textit{micro-grid} -- a self-contained region within a city-scale power grid -- such demand forecasting is crucial to plan reliable supply of electricity to meet the demand through supply from the grid, purchase agreement, intermittent renewables, and even from EV storage. Making these predictions locally within a micro-grid is important since the time to respond to irregularities is low and the latency for decision making is small within a microgrid given its lower inertia relative to the city grid~\cite{joshi2023survey}.

\begin{figure}[t]
    \centering
    \includegraphics[width = \columnwidth, trim=4cm 0cm 4.5cm 0.7cm, clip]{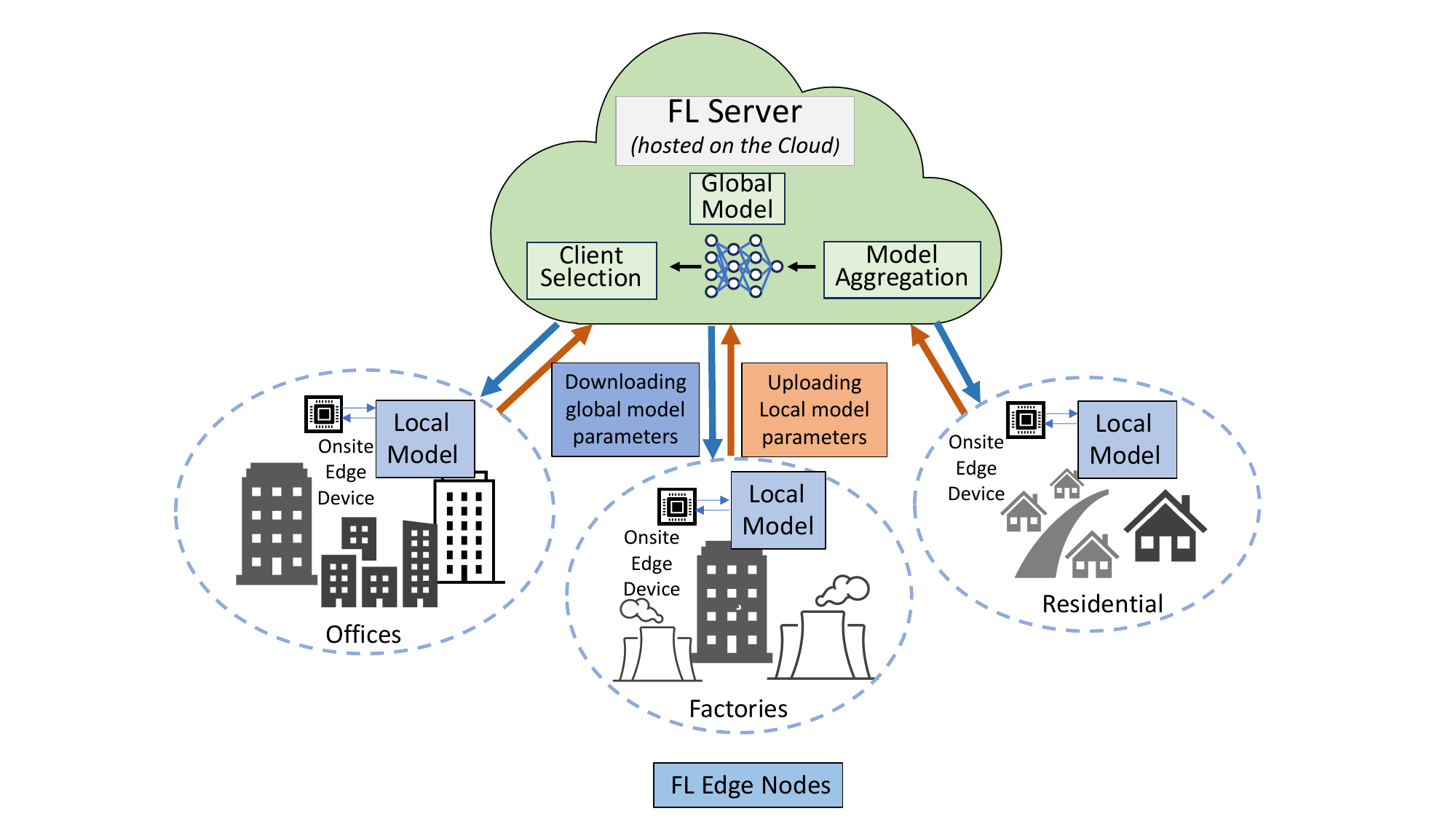}

    \caption{FL training in a power grid across edge and cloud. Circles indicate clusters of similar consumers.}
    \label{fig:FL_arch_ref}
\end{figure}

\subsection{Motivation}
Recently, there has been heightened privacy concerns and regulatory requirements that restrict central sharing of consumer-level consumption data at fine temporal scales to a city utility or micro-grid provider hosted on the cloud. These can expose the pattern of activities for the residential or commercial consumer, which in turn affects their privacy (e.g., when someone leaves for work and returns), and even expose proprietary enterprise information (e.g., power spike before a major release). 

\textit{Federated Learning (FL)} is a collaborative Machine Learning (ML) approach that allows decentralized edge devices or clients to train a global model on the cloud without sharing their local data centrally~\cite{google-FL}. In a FL setup (Fig.~\ref{fig:FL_arch_ref}), each energy consumer autonomously trains a forecasting model on their local historical time-series data on an edge device, shares the trained local model with the micro-grid provider or city utility hosted in the cloud, have these be aggregated in the cloud server to create a global forecasting model, and perform this over multiple iterations till the global model converges. The global demand prediction model thus trained can make accurate predictions for \textit{all entities} within the micro-grid. This ensures data privacy for each entity in the micro-grid, while allowing the benefit of a more accurate and collaborative forecasting model to be shared across all grid users.

\subsection{Gaps}
While short-term load forecasting using machine learning has been extensively studied -- employing techniques such as regression models, deep neural networks (DNNs)~\cite{duttagupta2023exploring}, ARIMA~\cite{alberg2018short}, and support vector machines~(SVMs)~\cite{nie2012hybrid} -- such approaches assume centralized training. Studies like \cite{Edwards2012PredictingFH} have demonstrated that DNNs often outperform traditional models, especially in commercial building settings. However, the application of these models using FL for microgrid-level power prediction remains under-explored. In particular, there is a lack of a principled understanding of how factors such as model architecture, loss functions, and non-IID data distributions affect the convergence and performance of the global model in FL. Furthermore, scaling FL to a large population of clients introduces challenges due to limited edge compute, communication bandwidth and non-IID data skews on clients. Further, there exists limited empirical evidence on how FL performs in heterogeneous microgrid environments with diverse load profiles, as well as a lack of optimization techniques tailored for FL that balance forecasting accuracy with communication and resource efficiency. These gaps motivate the need for a systematic investigation into scalable and effective FL frameworks for power forecasting in microgrids.

\subsection{Contributions}
In this paper, we study the benefits of clustering techniques and exponentially weighted loss for training RNN-based networks for commercial demand forecasting. The clustering strategy groups consumers with similar consumption patterns, and employing Federated Learning (FL) individually within each cluster mitigates data heterogeneity, reducing client-drift and improving forecasting accuracy. We use an exponentially weighted loss function, which facilitates a unified model design capable of forecasting demand across multiple time windows, effectively integrating recent data contributions while maintaining model stability and accuracy. We then combine the learnings from these studies to propose optimizations over FedAvg~\cite{mcmahan2023communicationefficient}, which achieves a faster convergence and a more generalized global model.

We make the following specific contributions:

\begin{enumerate}
\item We adopt a principled approach to evaluate the efficacy of clustering techniques and exponentially weighted loss functions in a FL setup, with the goal of training a global model for power demand prediction that generalizes well across a diverse population of consumers. Specifically, we perform ablation studies to assess the impact of using standard Mean Squared Error (MSE) and exponentially weighted MSE loss functions, a clustering-based client grouping strategy to address data heterogeneity, and different time-series forecasting architectures, including LSTM and GRU models.

\item Building on insights from our ablation study, we propose a set of optimizations to FedAvg~\cite{mcmahan2023communicationefficient}, one of the most widely used and foundational FL algorithms, tailored for demand forecasting in commercial microgrids and utility networks with heterogeneous consumers. Our training is performed using only univariate demand profiles -- 15-minute interval energy consumption data (in kWh) -- without incorporating any exogenous features such as weather or calendar information. This design enables us to evaluate the generalization capabilities of FL in a constrained, yet realistic, setting. Our approach strikes a balance between two extremes: highly customized models such as ARIMA trained individually for each consumer, and a single centralized model trained on aggregated data. Both of these are used as baselines to benchmark the effectiveness of our proposed federated strategy.

\item Finally, we conduct a comprehensive evaluation of our methodology using the OpenEIA~\cite{OpenEIA} dataset, focusing on three geographically and functionally diverse U.S. states -- California, Florida, and Rhode Island. Our results demonstrate the scalability and effectiveness of the proposed federated learning approach, achieving forecasting accuracies of up to 91\% while training on data from only 120 consumers and generalizing to over 30,000 consumers. We also validate this by training the FL model on a Raspberry Pi edge cluster. This highlights the ability of our method to provide accurate and scalable demand forecasts across heterogeneous populations with minimal training data.
\end{enumerate} 

\section{Related Work}
\label{sec:related-work}

\subsection{Time-series Forecasting Models}
\label{subsec:related-work-time-series-forecasting-models}
Time-series forecasting for energy demand has been widely studied using both traditional statistical methods and modern deep learning approaches. Perifanis et al.\cite{Perifanis_2023} compare several time-series forecasting models, including Multi-Layer Perceptrons (MLP), Convolutional Neural Networks (CNN), Recurrent Neural Networks (RNN), Long Short-Term Memory (LSTM), and Gated Recurrent Units (GRU), evaluating their performance in both centralized and federated training settings. Their study finds that LSTM and GRU consistently outperform MLP and CNN, which tend to exhibit greater instability across both validation and test datasets. Among the recurrent models, LSTM is often preferred over GRU due to its higher robustness to noise while achieving comparable forecasting accuracy~\cite{arunkumar2022comparative}. In addition to neural forecasting models, traditional statistical methods such as ARIMA have been widely used for short-term load forecasting in microgrid contexts~\cite{lee2011short}, and we include a variant of ARIMA as one of our baselines. There are also various accuracy metrics that have been proposed in the context of energy consumption predictions for smart grid applications~\cite{aman2014holistic}.

\subsection{Federated Learning}
\label{subsec:related-work-federated-learning}
Federated Learning (FL) was first introduced by Google to preserve user data privacy while training models on devices such as smartphones for applications like Gboard~\cite{mcmahan2023communicationefficient}. FL follows a client-server architecture in which data remains local to each client. Instead of uploading data to a central server, clients train models locally on their data and periodically share only model updates with the server, where these updates are aggregated to form a global model. This approach not only preserves privacy but also reduces communication overheads.

The foundational FL algorithm, \textit{FedAvg}\cite{mcmahan2023communicationefficient}, performs a weighted averaging of local model updates received from clients in each training round. Since its introduction, several extensions have been proposed to address practical challenges in real-world deployments. These challenges are commonly categorized into three broad types: \textit{data heterogeneity}, where clients possess non-IID data with varying volumes\cite{wolfrath2022haccs}; \textit{device heterogeneity}, where clients differ in computational capabilities, often varying across rounds~\cite{chai2020tifl}; and \textit{behavioral heterogeneity}, where network conditions and client availability fluctuate over time~\cite{abdelmoniem2023refl}.

There are also several frameworks for performing federated learning, including the popular Flower framework~\cite{flower}, Flotilla to compose modular and scalable FL strategies~\cite{banerjee2025flotilla}, and Flox for FL on the edge using Functions as a Service~\cite{kotsehub2022flox}.

\subsection{Federated Power Forecasting}
FL has been used in diverse domains, ranging from healthcare~\cite{singh2022framework} and crowd-sourced weather prediction~\cite{de2024federated} to collaborating using Jupyter notebooks~\cite{krishnasamy2024collaborative}.
The rapid proliferation of smart meters equipped with local computing capabilities has paved the way for practical implementations of FL, driving research into efficient and privacy-preserving load forecasting techniques. Recent studies have tackled key challenges such as the limited computational resources of smart meters~\cite{Savi2021ShortTermEC}, data privacy and security concerns~\cite{sievers2023secure}, and the inherently non-IID nature of energy consumption data across diverse consumers. These efforts have yielded promising results for time-series load forecasting using FL on edge devices, especially in commercial and residential settings~\cite{taik2020electrical}.

Concurrently, as microgrid architectures become more complex, cloud-based energy management systems (EMS) have been widely adopted for real-time monitoring and forecasting of both energy production and consumption~\cite{rosero2021cloud}. Recent work~\cite{savi-short-term} highlights the effectiveness of hybrid edge–cloud architectures in enabling federated training of load forecasting models within commercial environments.

Building on this paradigm, our proposed methodology leverages smart meters at the edge to perform local model updates while employing cloud infrastructure for orchestration and aggregation. This approach strikes a balance between privacy, scalability, and performance. Furthermore, our work advances prior research by rigorously investigating the key factors influencing the quality and generalizability of federated forecasting models in smart grid settings. Specifically, we explore lightweight model architectures, custom loss functions, and client clustering strategies to achieve low-latency, high-accuracy predictions suitable for large-scale deployment.

\section{Methodology}
\label{sec:methodology}
To address the challenges posed by heterogeneous commercial power consumption data in federated learning, we propose a comprehensive methodology that combines three key components: (1) \textit{consumer clustering} to group clients with similar consumption patterns, (2) the use of \textit{custom loss functions}, namely an exponentially weighted MSE loss, to better capture temporal dynamics during training, and (3) evaluation of different \textit{time-series forecasting architectures}, specifically Long Short-Term Memory (LSTM) and Gated Recurrent Unit (GRU) models, to identify optimal model structures for federated forecasting.

In the following subsections, we detail each of these components and describe how they are integrated within our federated learning framework.

\subsection{K-means Clustering}
\label{subsec:clustering}

In FL, client-level heterogeneity -- especially in data distributions -- can significantly degrade the performance and convergence of the global model. This is particularly true in power demand forecasting, where consumers exhibit widely varying usage patterns due to differences in building type, occupancy, operational hours, and appliances. Clustering groups consumers with similar time-series consumption patterns enables the identification of shared trends and seasonality within subsets of the data. This approach addresses data heterogeneity at a coarse level and improves forecasting accuracy by tailoring models to the specific characteristics of each cluster. 

Fig.~\ref{fig:box plot for California State(mean consumption KWH)} shows the distribution of mean 15-minute electricity consumption across consumers from the OpenEIA dataset~(detailed in \S~\ref{subsec:dataset}) for three US states: California, Florida, and Rhode Island. The majority of consumers exhibit low average utilization, while a smaller subset demonstrates significantly higher consumption levels, indicating long-tailed distribution where most clients follow low-demand patterns, but a few contribute disproportionately to overall consumption. By clustering consumers according to their consumption profiles and training separate models for each cluster, we can build more generalized and robust forecasting models.

We use K-means clustering to partition consumers into distinct clusters based on the similarities in their data. Each cluster represents a different behavior or state in the time series. After clustering, we train separate forecasting models (i.e., LSTM, GRU) for each cluster in a federated manner using the FedAvg algorithm for aggregation, incorporating the unique characteristics of the data in each segment into the global model.

We estimate the daily average of the energy consumption (kWh) time series and obtain a vector for each building across the clustering duration (e.g., in our experiments, a vector of size 273 with mean daily kWh values for approximately 9 months). These vectors are used for the K-means clustering. Determining the optimal number of clusters (K) is a key consideration. We use the elbow method and silhouette score to help find an appropriate K value. We train a model for each cluster in a federated manner.

\subsection{Forecasting Models}
\label{subsec:models}

From the popular time-series DNN model architectures, we consider LSTM and GRU:
\subsubsection{LSTM model}
LSTM networks are a form of Recurrent Neural Network~(RNN) architecture designed to address the vanishing gradient problem encountered by traditional RNNs. LSTMs are particularly effective in capturing long-range dependencies in sequential data and are, thus, widely used in various tasks such as natural language processing, speech recognition, and time series forecasting.

The core component of LSTM is the \textit{memory cell}, which maintains a hidden state (\( c_t \)) that can store information over long periods of time. The \textit{forget gate (\( f_t \))} controls what information from the previous cell state should be discarded or kept. It takes as input the previous hidden state (\( h_{t-1} \)) and the current input (\( x_t \)) and outputs a value between 0 and 1 for each component of the cell state, indicating how much of the corresponding information to forget.
The \textit{input gate (\( i_t \))} determines what new information should be stored in the cell state. It similarly takes the previous hidden state (\( h_{t-1} \)) and the current input (\( x_t \)) and decides which values to update.
An \textit{update gate (\( g_t \))} computes the updates to the cell state based on the previous hidden state (\( h_{t-1} \)) and the current input (\( x_t \)). This gate controls the extent to which the cell state (\( c_{t-1} \)) is updated.
Finally, the \textit{output gate (\( o_t \))} controls what information should be output as the hidden state (\( h_t \)). It considers the previous hidden state (\( h_{t-1} \)) and the current input (\( x_t \)), and it modulates the information in the cell state to produce the next hidden state.

The computations within an LSTM unit can be summarized with the following equations:
\begin{align*}
    f_t &= \sigma(W_f \cdot [h_{t-1}, x_t] + b_f) \\
    i_t &= \sigma(W_i \cdot [h_{t-1}, x_t] + b_i) \\
    g_t &= \text{tanh}(W_g \cdot [h_{t-1}, x_t] + b_g) \\
    c_t &= f_t \odot c_{t-1} + i_t \odot g_t \\
    o_t &= \sigma(W_o \cdot [h_{t-1}, x_t] + b_o) \\
    h_t &= o_t \odot \text{tanh}(c_t)
\end{align*}
Here, \( \sigma \) represents the sigmoid activation function, \( \odot \) denotes element-wise multiplication, \( \text{tanh} \) is the hyperbolic tangent function, and \( W \) and \( b \) are weight matrices and bias vectors, respectively.

This gating mechanism addresses the vanishing gradient problem, making LSTMs effective for tasks involving sequential data processing, such as time series forecasting. LSTMs learn to capture both short-term and long-term dependencies in data, making them versatile for applications where understanding context and temporal patterns is crucial.

\subsubsection{GRU model}
Gated Recurrent Units~(GRUs) are also a type of RNN designed to address the challenges of capturing long-term dependencies in sequential data while mitigating issues like the vanishing gradient problem. GRUs incorporate two main gates: the update gate ($z$) and the reset gate ($r$), which regulate the flow of information throughout the network. These gates enable GRUs to selectively retain or discard information from previous time steps. 

Mathematically, the \textit{update gate ($z_t$)} at time step $t$ is computed using the sigmoid activation function $\sigma$ as follows:
\[ z_t = \sigma(W_z \cdot [h_{t-1}, x_t]) \]
where $W_z$ represents the weight matrix for the update gate, and $[h_{t-1}, x_t]$ denotes the concatenation of the previous hidden state $h_{t-1}$ and the current input $x_t$. Similarly, the \textit{reset gate ($r_t$)} is computed in a similar manner:
\[ r_t = \sigma(W_r \cdot [h_{t-1}, x_t]) \]
The candidate \textit{activation ($\tilde{h}_t$)}, which proposes the new memory content, is calculated using the current input $x_t$, the previous hidden state $h_{t-1}$, and the reset gate ($r_t$):
\[ \tilde{h}_t = \text{tanh}(W_h \cdot [r_t \odot h_{t-1}, x_t]) \]
Here, $\odot$ denotes element-wise multiplication, and $W_h$ is the weight matrix for the candidate activation. Finally, the new memory content ($h_t$) is computed by combining the previous memory content weighted by the update gate and the candidate activation weighted by $(1 - z_t)$:
\[ h_t = z_t \odot h_{t-1} + (1 - z_t) \odot \tilde{h}_t \]

Both LSTM and GRU architectures are effective for modeling temporal patterns in sequential energy consumption data, with GRUs providing a lightweight but less accurate model compared to LSTMs. We evaluate both models in the context of federated load forecasting to understand their relative performance and suitability for deployment on resource-constrained edge devices.

\subsection{Custom Loss Function for Forecasting Model Training}
\label{subsec:loss-func}

Effective training of forecasting models in a federated setting requires carefully chosen loss functions that align with the temporal dynamics of the target variable. In this work, we explore two loss functions for training our recurrent models -- standard MSE and a modified Exponentially Weighted MSE (EW-MSE).

\subsubsection{MSE}
The standard Mean Squared Error (MSE) is widely used for regression tasks, including time series forecasting. It measures the average squared difference between the predicted and actual values across all forecasted time steps:

\[
\text{MSE}(y, \hat{y}) = \frac{1}{N} \sum_{i=1}^{N} (y_i - \hat{y}_i)^2
\]
where \( y_i \) and \( \hat{y}_i \) denote the ground truth and predicted values at the \( i^{\text{th}} \) time step, respectively, and \( N \) is the prediction horizon. While MSE provides a straightforward and symmetric measure of error, it treats all time steps uniformly. However, in recurrent models such as LSTMs and GRUs, prediction errors typically accumulate over time due to the compounding of inaccuracies at each step. Consequently, later time steps tend to exhibit higher error variance.

\subsubsection{EW-MSE}
To account for the increased difficulty of making accurate predictions at further horizons, we introduce an Exponentially Weighted Mean Squared Error (EW-MSE) loss function. EW-MSE emphasizes accuracy on longer-range predictions by assigning exponentially increasing weights to errors at later time steps:

\[
\text{EW-MSE}(y, \hat{y}) = \frac{1}{N} \sum_{i=1}^{N} \beta^{i-1} \cdot (y_i - \hat{y}_i)^2
\]

Here, \( \beta > 1 \) is a tunable hyperparameter that controls the degree of emphasis placed on future predictions. When \( \beta = 1 \), EW-MSE reduces to the standard MSE. By amplifying the penalty for errors further ahead in the prediction horizon, EW-MSE encourages the model to more carefully learn the dynamics governing long-range behavior -- an aspect particularly beneficial in energy forecasting scenarios where cumulative errors can severely degrade performance.

As we demonstrate in \S~\ref{sec:Exponential weighed MSE Loss}, using EW-MSE in local model training leads to improved generalization and overall forecasting accuracy at the global level in our federated learning setup.

\subsection{Proposed Federated Learning Algorithm}

\begin{algorithm}[t!]
\footnotesize
    \SetAlgoNlRelativeSize{-1}
    \SetAlgoNlRelativeSize{-1}
    \SetKwInOut{Input}{Input}
    \SetKwInOut{Output}{Output}
    
    \caption{Federated Learning Scheme}
    \Input{number of all clients $N$; number of randomly selected clients per round $M$; learning rate $\eta$; local minibatch size $B$; number of local epochs in each communication $E$; total number of communication round $T$; initial global model parameters $w_0$; set of clusters $C$;period for daily averaged energy consumption $t_p$}
    \KwOut{updated cluster parameters $w^{c_k}$ for $k^{th}$ cluster }
    
    \SetKwFunction{FederatedTraining}{FEDERATED\_TRAINING}
    \SetKwFunction{ConsumptionSummary}{ConsumptionSummary}
    \SetKwFunction{Clustering}{Clustering}
    \SetKwFunction{ClientUpdate}{ClientUpdate}
    
    \SetKwProg{Fn}{Function}{:}{}
    
    \Fn{\FederatedTraining{$N, M, \eta, B, E, T, w_0,t_p$}}{
        \For{$n = 1, 2, \ldots, N$}{
    $ z_n  \leftarrow \ConsumptionSummary(t_p) $\;
    $Z \leftarrow \{z_1, z_2, \ldots, z_N\}$\;
    ${c_1, c_2, \ldots, c_k} \leftarrow \text{Clustering}(Z)$\;
    $C \leftarrow \{{c_1, c_2, \ldots, c_k}\}$\;
}
    \For{all $c_k \in C$}{
        Initialize $w^{c_k} \leftarrow w_{0}$\;
    }               
        
        \For{each cluster $c_k$ in $C$ , in parallel}{
    $w_{t}^{c_k} \leftarrow w^{c_k}$\;
    \For{$t = 1, 2, \ldots, T$}{
        $s_t \leftarrow$ Server randomly selects $M$ clients out of all clients in $c_k$\; 
        \For{each client $c^i_k$ in $s_t$ in parallel}{
            $w_{t+1}^{c_k^{i}} \leftarrow$ \texttt{ClientUpdate}$(c^i, w_{t}^{c_k})$\;
        }
        $w^{c_k}_{t+1} \leftarrow \frac{1}{\left|s_t\right|} \sum_{i \in s_t } w_{t+1}^{c_{k}^{i}}$\;
    }
}}

    \Fn{\ClientUpdate{$c, w$}}{
        Create mini-batches of client $c$ into batches of size $B$\;
        \For{each local training epoch $e = 1, 2, \ldots, E$}{
            \For{each minibatch $B$}{
                $w \leftarrow w - \eta \cdot \nabla l_{ew-mse}(w; b)$\;
            }
        }
        \Return $w$ to the server
    }
    \Fn{\ConsumptionSummary{$t_p$}}{
        $z$ =  daily averaged energy consumption historic data  for a  period $t_p$\;
        \Return $z$ 
    }
\label{alg:fl_scheme}

\end{algorithm}

Algorithm~\ref{alg:fl_scheme} provides an overview of our federated learning (FL) algorithm. As an optional, one-time pre-processing step, we employ a \textit{clustering scheme} that is performed centrally in a privacy-preserving manner. Specifically, participating clients upload their consumption summary vectors \( z_k \), which consist of \textit{daily averaged energy consumption data} over a defined period \( t_p \). This process involves coarsening the original 15-minute granularity data into daily 24-hour averages, effectively reducing data resolution and thereby helping to preserve client privacy by limiting exposure of fine-grained usage patterns.  These summaries are then clustered and clients are assigned to their respective cluster index $c_k \in C$. The FL training then happens on each cluster independently. We evaluate clustering using $K$-means clustering (\S~\ref{subsec:clustering}), and a default (non-clustered) approach, training a single global model.

For each cluster, the FL server initializes the model weights $w_0$ for the DNN architecture, e.g., LSTM (\S~\ref{subsec:models}), and distributes them to the edge clients from that cluster participating in a round. We select a random subset of $n$ clients within a cluster, where $n$ is configurable.

The clients train the DNN model received from the cloud using their local consumption data for the specified number of epochs using the relevant loss function~(\S~\ref{subsec:loss-func}).
The EW-MSE loss function we propose is flexible enough to be used by the clients for local training irrespective of the cluster they belong to. Each edge-client then uploads the trained model to the cloud server.
The server uses synchronous aggregation where it waits for all clients to return their local models, and uses FedAvg to aggregate the local models into a new global model. This is one round of training, and is repeated for $T$ rounds, with the updated global model used to bootstrap local training in a new round.

\section{Experimental Setup}
\label{sec:Experimental Setup}

\subsection{Dataset used in Experiments and Analysis}
\label{subsec:dataset}

We use dataset from US Department of Energy's \textit{Open Energy Data Initiative (Open EIA)}~\cite{OpenEIA}, which provides energy consumption of residential and commercial buildings in USA.

\begin{figure}[t!]
\centering
\includegraphics[width=0.7\columnwidth]{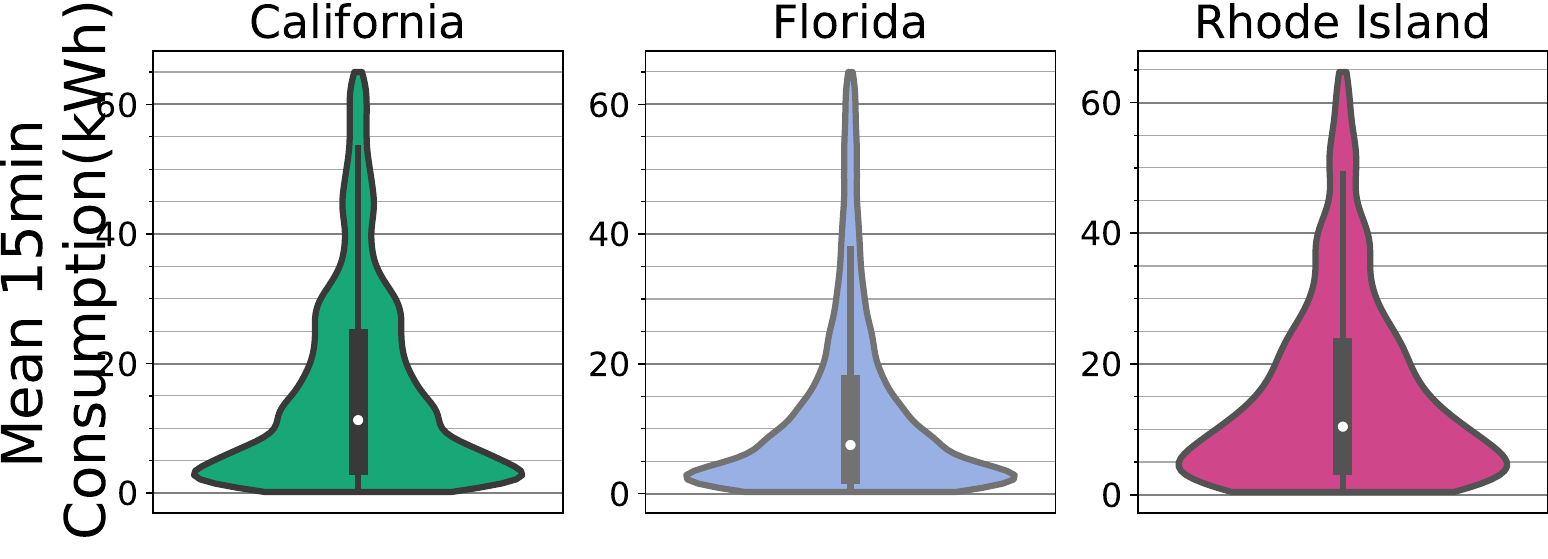}
    \caption{Violin Plots for mean consumption over 15 minute kWh data for commercial buildings in CA, FLO and RI, as obtained from the\textit{ Open EIA dataset}.}
    \label{fig:box plot for California State(mean consumption KWH)}
    \vspace{-0.1in}
 \end{figure}

The dataset (Table~\ref{tab:datasets}) includes a time series of energy usage data, measured in kWh, for dwelling units in commercial/residential buildings, organized by the state in which the building is located and the year of measurement. For the purpose of this paper, we use the \textit{2023 release 1} for \textit{comstock}, which reports data for a subset of commercial buildings in a state-wise manner. The data is reported at 15~min granularity, and 1 year of data is available. We choose data for \textit{commercial buildings} in the states of \textit{California~(CA)}, \textit{Florida~(FLO)} and \textit{Rhode Island~(RI)}. Fig.~\ref{fig:box plot for California State(mean consumption KWH)} shows the distribution of their mean energy consumption~(in kWh). The minimum consumption is at $0.16$~kWh, the maximum at $63.8$~kWh, median at $12.7$~kWh, and Q1 and Q3 are at $4.7$~kWh and $28.4$~kWh. Around $3200$ buildings have mean consumption greater than $63.8$~kWh.

\paragraph*{Large Held-out Test Set}
\label{subsec:Held-out set testing}
We set aside a large fraction of held-out buildings, i.e., clients that are not used in training, for validating the predictions made using the trained global FL model. The held-out buildings for CA is $39,291$ out of $39,391$ available~(i.e., only 100 used for training). Similarly, FLO has $24,344$ and RI has $1276$ buildings in their held-out sets, with data from the same number~(100) of buildings from each of these datasets are used for training our model.

\begin{table}[t]
\centering
\footnotesize
\setlength{\tabcolsep}{1.5pt} 
\caption{Open EIA data used in our experiments}
\label{tab:datasets}

\begin{tabular}{L{1.75cm}|R{1.25cm}|R{1.25cm}|R{1.25cm}|R{1cm}|R{1.5cm}}
\hline
\bf State &
\bf \# of buildings &
\bf \# of train buildings &
\bf \# of test buildings &
\bf Data Year &
\bf Samples per Building \\ \hline\hline
California (CA)   & 39,391 & 100 & 39,291 & 2018 & 35,040 \\ \hline
Florida (FLO)     & 24,444 & 100 & 24,344 & 2018 & 35,040 \\ \hline
Rhode Island (RI) & 1376  & 100 & 1376  & 2018 & 35,040 \\ \hline
\end{tabular}%
\end{table}

\subsection{Data Preparation and System Configuration}

In our experiments, each building functions as an independent FL client, resulting in a total of 100 clients whose consumption data remain private to each participant. Each client trains a local LSTM or GRU model on its own 15‑minute interval usage data, which is first normalized to the \([0,1]\) range using Min–Max scaling over the entire year. We frame the forecasting task using a look‑back window of 8 steps (2 hours) and a look‑ahead horizon of 4 steps (1 hour). The resulting time series for each client is partitioned into training and testing sets in a 75:25 ratio -- approximately the first nine months of data for training and the final three months for evaluation. The federated training proceeds for 500 rounds, with each client performing local updates before the server aggregates model parameters according to our proposed FL scheme.  

\subsection{Baseline}

As a baseline, we compare our proposed approach against a \textit{Seasonal ARIMA (SARIMA)} model, which is a standard technique for forecasting time series with pronounced seasonality. Methods based on Autoregressive Integrated Moving Average (ARIMA) models are widely used for time series forecasting, and also for demand forecasting in power grids. We use the auto-ARIMA implementation that is part of the \texttt{pmdarima}~(Pyramid ARIMA) Python library, which automates the selection of optimal parameters for training the SARIMA models. We initially train our SARIMA model over a 30-day period. After making predictions for a specific duration, we periodically (every 30 days) recalibrate the model by re-fitting the model's parameters, improving the model's performance. 

\subsection{K-means Clustering Hyperparameters}
\label{subsec:Kmeans Clustering}

We set $k=4$ clusters, as observed to be the optimal number of clusters from an elbow plot, and sample 25 buildings per cluster for training across 100 buildings per state. The client consumption vector used for clustering $t_p$ is the 273-dimensional daily average consumption~($\approx 9$ months) per building.
We train 5 different flavors of models. While $F^{A}$ is the FedAvg FL model trained on all 100 buildings, $F^{C1}, F^{C2}, F^{C3} \text{ and } F^{C4}$~(see Table~\ref{tab:120_buildings_accuracy_K-means_comparison} and Table~\ref{tab:Table 10}) are the models for each of the 4 clusters, each trained on 25 buildings. We validate against a large held-out set per state, and on a smaller set of 30 buildings on a per-cluster basis.

\subsection{Evaluation Metrics}
\label{subsec:Metrics}

To evaluate the quality of our federated load forecasting models, we use three complementary metrics -- Root Mean Squared Error (RMSE), Mean Absolute Percentage Error (MAPE), and Accuracy -- each tailored to highlight different aspects of prediction performance.

\subsubsection{Root Mean Squared Error (RMSE)} 
RMSE measures the square root of the average squared differences between the actual and predicted consumption values. In the context of power demand forecasting, RMSE is particularly useful for penalizing large deviations, ensuring that models which occasionally produce significant errors are correctly identified and improved upon:
\[
\mathrm{RMSE} = \sqrt{\frac{1}{n} \sum_{i=1}^{n} \bigl(Y_{\text{actual},i} - Y_{\text{predicted},i}\bigr)^2}.
\]

\subsubsection{Mean Absolute Percentage Error (MAPE)} 
MAPE expresses the average absolute error as a percentage of the actual consumption, providing a scale‐independent measure that facilitates comparison across buildings with different load profiles. This normalization is critical when evaluating models over heterogeneous consumers:
\[
\mathrm{MAPE} = \frac{1}{n} \sum_{i=1}^{n} \left|\frac{Y_{\text{actual},i} - Y_{\text{predicted},i}}{Y_{\text{actual},i}}\right|\times 100\%.
\]


\begin{table*}[t!]
\centering
\caption{\textit{Accuracy}~(in \%) for cluster specific models $F^{C1}$, $F^{C2}$, $F^{C3}$, $F^{C4}$ and single FedAvg $F^{A}$ model. 60~min forward predictions on a large held-out set of test buildings of CA. Here we use an LSTM model and MSE loss.}
\label{tab:Table 10}
\begin{tabularx}{\textwidth}{l| >{\RaggedLeft\arraybackslash}X | >{\RaggedLeft\arraybackslash}X | >{\RaggedLeft\arraybackslash}X | >{\RaggedLeft\arraybackslash}X | >{\RaggedLeft\arraybackslash}X | >{\RaggedLeft\arraybackslash}X}
\toprule
& \textbf{$F^{A}$} \textbf{model} & \textbf{$F^{C1}$} \textbf{model} & \textbf{$F^{C2}$} \textbf{model} & \textbf{$F^{C3}$} \textbf{model} & \textbf{$F^{C4}$} \textbf{model} & \textbf{\begin{tabular}[c]{@{}l@{}}Avg of\\ Averages\\($F^{C1}:F^{C4}$)\end{tabular}} \\
\midrule
\midrule
\textbf{Cluster 1} & 88.0 & \cellcolor[HTML]{34FF34}\textbf{88.8} & - & - & - & - \\\hline
\textbf{Cluster 2} & 91.1 & - & \cellcolor[HTML]{34FF34}\textbf{91.4} & - & - & - \\\hline
\textbf{Cluster 3} & 85.0 & - & - & \cellcolor[HTML]{34FF34}\textbf{84.2} & - & - \\\hline
\textbf{Cluster 4} & 90.3 & - & - & - & \cellcolor[HTML]{34FF34}\textbf{91.3} & - \\\hline
\textbf{Average} & 88.6 & 88.8 & 91.4 & 84.2 & 91.3 & \cellcolor[HTML]{34FF34}\textbf{88.9} \\
\bottomrule
\end{tabularx}
\end{table*}

\subsubsection{Accuracy} 
We define Accuracy as the complement of MAPE, yielding an intuitive percentage score that reflects how closely, on average, the forecasts match the true consumption values. High Accuracy indicates consistently reliable predictions across all clients, where \(\mathrm{Accuracy} = 100\% - \mathrm{MAPE}\). By jointly considering RMSE, MAPE, and Accuracy, we obtain a holistic view of forecasting performance, capturing both absolute error magnitudes and relative, scale‐adjusted accuracy across our set of 100 federated clients.

\begin{table}[t]
\centering
\caption{\textit{Average Accuracy}~(in \%) for global LSTM model $F^{A
}$ and cluster-specific LSTM models $F^{C1},F^{C2},F^{C3},F
^{C4}$ trained with MSE loss and SARIMA models $S^{C1},S^{C2},S^{C3},S^{C4}$. 60~min forward prediction for 4~clusters of 30~buildings of CA. 
FL-LSTM models are competitive with the SARIMA model despite not requiring retraining.}
\label{tab:120_buildings_accuracy_K-means_comparison}

\begin{tabular}{l|l|l|l|l}
\hline
            & \textbf{Cluster 1}        & \textbf{Cluster 2}        & \textbf{Cluster 3}        & \textbf{Cluster 4}        \\ \hline \hline
\textbf{$\mathbf{F^A}$ model} & \multicolumn{1}{r|}{90.9} & \multicolumn{1}{r|}{93.6} & \multicolumn{1}{r|}{87.5} & \multicolumn{1}{r}{91.6} \\
\midrule
\textbf{$\mathbf{F^{C1}}$ model} & \multicolumn{1}{r|}{91.2}          & -                                   & -                                   & -                                   \\ \hline
\textbf{$\mathbf{S^{C1}}$ model} & \multicolumn{1}{r|}{\textbf{\cellcolor[HTML]{34FF34}{93.2}}} & -                                   & -                                   & -                                   \\
\midrule
\textbf{$\mathbf{F^{C2}}$ model} & -                                  & \multicolumn{1}{r|}{\textbf{\cellcolor[HTML]{34FF34}{93.7}}} & -                                   & -                                   \\ \hline
\textbf{$\mathbf{S^{C2}}$ model} & -                                  & \multicolumn{1}{r|}{92.6}          & -                                   & -                                   \\ \midrule
\textbf{$\mathbf{F^{C3}}$ model} & -                                  & -                                   & \multicolumn{1}{r|}{88.8}           & -                                   \\ \hline
\textbf{$\mathbf{S^{C3}}$ model} & -                                  & -                                   & \multicolumn{1}{r|}{\textbf{\cellcolor[HTML]{34FF34}{89.1}}} & -                                   \\
\midrule
\textbf{$\mathbf{F^{C4}}$ model} & -                                  & -                                   & -                                   & \multicolumn{1}{r}{91.6}          \\ \hline
\textbf{$\mathbf{S^{C4}}$ model} & -                                  & -                                   & -                                   & \multicolumn{1}{r}{\cellcolor[HTML]{34FF34}{\textbf{91.8}}} \\ \hline
\end{tabular}%
\end{table}

\section{Results}
In this section, we present the results of the evaluation of each of our proposed optimizations. 

\subsection{Impact of Clustering}
\label{subsec:restults-clustering}

We first evaluate the impact of clustering on federated LSTM training by applying \(K\)-means with \(k=4\) to partition 120 California buildings into four clusters of 25 clients each. Each cluster trains its own LSTM model using the standard MSE loss, and we compare these cluster‑specific federated models \(F^{C1}\dots F^{C4}\) against two baselines: (a) SARIMA models fitted separately on each cluster (\(S^{C1}\dots S^{C4}\)), and (b) a single global FedAvg LSTM model \(F^A\) trained on all 120 buildings. Table~\ref{tab:120_buildings_accuracy_K-means_comparison} reports the accuracy of each model on 30 held‑out buildings per cluster. Across all clusters, the federated LSTM models trained on clustered clients consistently outperform the global model \(F^A\), demonstrating the benefit of grouping similar consumption patterns. Moreover, the accuracies of \(F^{Ci}\) closely match those of the corresponding SARIMA baselines \(S^{Ci}\), with \(F^{C2}\) even surpassing \(S^{C2}\), despite requiring no periodic retraining. This highlights the efficiency and generalization advantages of our approach.

To further assess scalability, we validate all models on a much larger test set of 39,290 California buildings, each assigned to one of the four clusters (Table~\ref{tab:Table 10}). Even with this extensive held‑out population, cluster‑specific federated models maintain their superiority over the single global model in most clusters, yielding an average accuracy of 88.98\%, compared to 88.60\% for \(F^A\). These results confirm that \(K\)-means clustering combined with per‑cluster federated training can slightly but consistently enhance forecasting performance relative to a traditional, unclustered FedAvg approach.

\begin{table}[t]
\setlength{\tabcolsep}{3.5pt} 
\centering
\footnotesize
\caption{
\textit{Accuracy} for single FedAvg~($F^{A}$) model trained using \textit{MSE loss} vs. \textit{EW-MSE loss}, for every $15^{th}$-min of a 1~h forward prediction with LSTM model. For these experiments, no clustering is applied. EW-MSE loss is better for all states and intervals.}
\label{tab:MSEvsEWMSE_15thmin}
\begin{tabular}{|l|c|c|c|c|c|c|c|c|}
\hline
& \multicolumn{2}{c|}{\textbf{$15^{th}$min}} & \multicolumn{2}{c|}{\textbf{$30^{th}$min}} & \multicolumn{2}{c|}{\textbf{$45^{th}$min}} & \multicolumn{2}{c|}{\textbf{$60^{th}$min}} \\
\cline{2-9}
\textbf{State} & \textbf{MSE} & \textbf{EW-MSE} & \textbf{MSE} & \textbf{EW-MSE} & \textbf{MSE} & \textbf{EW-MSE} & \textbf{MSE} & \textbf{EW-MSE} \\
\hline
\textbf{CA} & 91.9 & \cellcolor[HTML]{34FF34}\textbf{93.6} & 89.6 & \cellcolor[HTML]{34FF34}\textbf{91.9} & 87.4 & \cellcolor[HTML]{34FF34}\textbf{89.9} & 85.1 & \cellcolor[HTML]{34FF34}\textbf{87.9} \\
\hline
\textbf{FLO} & 88.5 & \cellcolor[HTML]{34FF34}\textbf{89.4} & 84.4 & \cellcolor[HTML]{34FF34}\textbf{85.7} & 80.4 & \cellcolor[HTML]{34FF34}\textbf{81.3} & 76.0 & \cellcolor[HTML]{34FF34}\textbf{76.7} \\
\hline
\textbf{RI} & 88.0 & \cellcolor[HTML]{34FF34}\textbf{90.1} & 84.2 & \cellcolor[HTML]{34FF34}\textbf{87.8} & 80.3 & \cellcolor[HTML]{34FF34}\textbf{85.5} & 76.0 & \cellcolor[HTML]{34FF34}\textbf{81.9} \\
\hline
\end{tabular}
\end{table}

\subsection{Benefits of EW-MSE Loss}
\label{sec:Exponential weighed MSE Loss}

\begin{figure*}[t]
\centering
\begin{minipage}[t]{.36\textwidth}
  \centering
\includegraphics[width=1\textwidth]{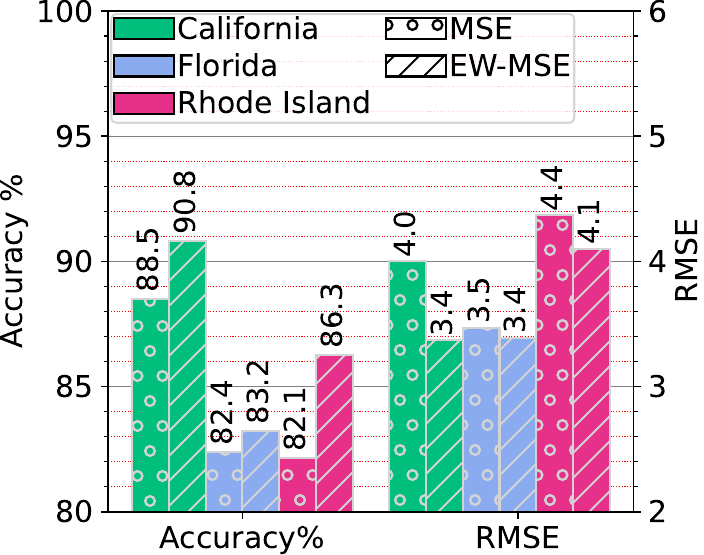}
    \caption{\textit{ Avg. Accuracy and RMSE} for models using \textit{MSE vs. EW-MSE loss} with LSTM model, without clustering.}
    \label{fig:MSEvsEWMSE}
\end{minipage}\hfill
\begin{minipage}[t]{.285\textwidth}
  \centering
\includegraphics[width=1\textwidth]{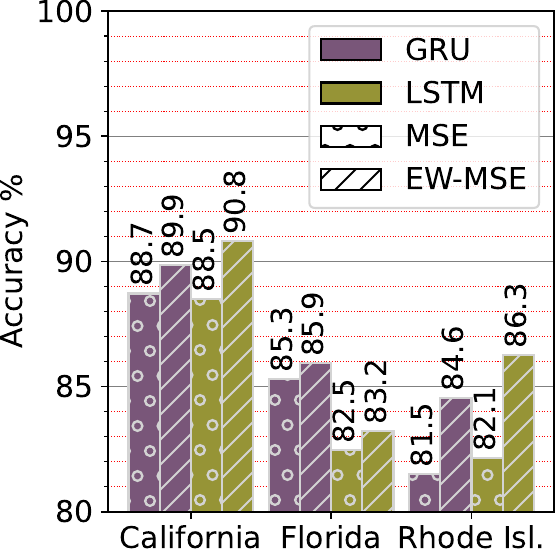}
    \caption{\textit{Avg. Accuracy of LSTM vs. GRU models} using \textit{MSE and EW-MSE losses}, without clustering.}
    \label{fig:GRUvsLSTM}
\end{minipage}\hfill
\begin{minipage}[t]{.3\textwidth}
  \centering
    \includegraphics[width=1\columnwidth]
    {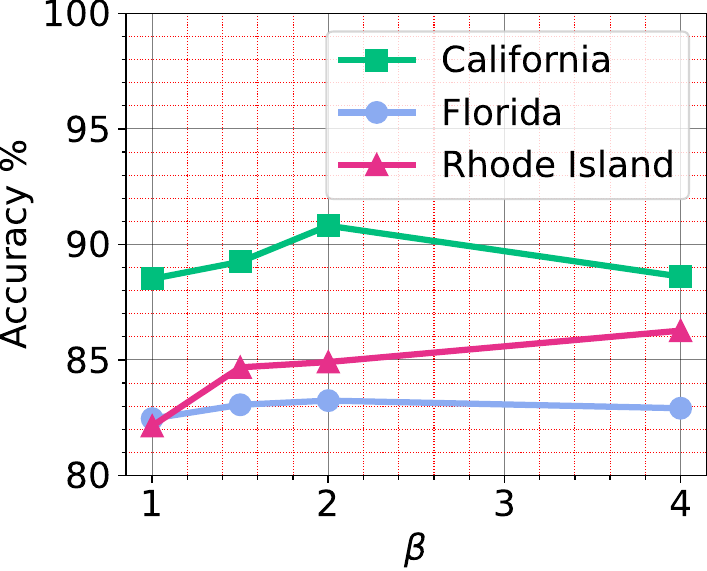}
    \caption{\textit{Avg. Accuracy} with different $\beta$ in EW-MSE loss with LSTM model.Here clustering is not applied.}
    \label{fig:beta ablations)}    
\end{minipage}
\vspace{-0.1in}
\end{figure*}

We evaluate the impact of the Exponentially Weighted MSE (EW‑MSE) loss by comparing it against the standard MSE loss for a single federated LSTM model across large held‑out datasets in California (CA), Florida (FL), and Rhode Island (RI). Table~\ref{tab:MSEvsEWMSE_15thmin} details the prediction accuracy at each 15‑minute forecasting horizon, while Fig.~\ref{fig:MSEvsEWMSE} summarizes the overall average accuracy and RMSE. As expected, the model trained with MSE exhibits declining accuracy for longer horizons; for example, in CA the 60‑minute prediction accuracy drops to 85.1\%. In contrast, EW‑MSE compensates for this degradation by penalizing errors at farther time points more heavily, yielding an average accuracy of 90.81\% in CA (vs.\ 88.15\% with MSE). Moreover, per‐interval improvements are observed across all states: in CA the 15‑minute horizon accuracy rises from 91.9\% to 93.6\% and the 60‑minute from 85.1\% to 87.9\%; in FL the 15‑minute accuracy increases from 88.5\% to 89.4\% and the 60‑minute from 76.0\% to 76.7\%; and in RI the 15‑minute accuracy climbs from 88.0\% to 90.1\% and the 60‑minute from 76.0\% to 81.9\%. These gains correspond to an average accuracy boost of approximately 1\% in FL and 4\% in RI. Together, these results demonstrate that EW‑MSE consistently enhances long‑horizon forecasting performance compared to vanilla MSE in our federated learning framework.

We also experiment with different $\beta$ values~($\beta\in[1,..,4]$), for EW-MSE loss, to understand the impact of this hyper-parameter. All values of $\beta > 1$ give a better performance in comparison with $\beta = 1$ (i.e MSE loss), with $\beta$ needing to be tuned for the best results.
Fig.~\ref{fig:beta ablations)} shows the result of this ablation for CA, FLO and RI.
This enhancement underscores the efficacy of incorporating exponential weighting to mitigate prediction challenges associated with longer time intervals in energy forecasting.

\subsection{Comparison of LSTM and GRU Models}
In Fig.~\ref{fig:GRUvsLSTM} we compare average accuracies of the large held-out set for the three states predicted using LSTM and GRU models trained using both MSE and EW-MSE losses. Results are reported for 39,290 test  buildings in CA, 12,344 test buildings in FLO and 1,276 test buildings in RI.
Both the GRU and LSTM models' performance is almost the same except for FLO, with GRU we see an improvement of $\approx3\%$ when using MSE loss and an improvement of $\approx2.5\%$ with EW-MSE loss. LSTM benefits more from the EW-MSE loss compared to GRU model as can be seen with an improvement of $\approx4\%$ with EW-MSE loss over MSE loss for RI with LSTM model in comparison to the $\approx3\%$ improvement with EW-MSE loss over MSE loss with GRU model,a similar trend can be observed for CA where an improvement of $\approx2.3\%$ with EW-MSE loss over MSE loss with LSTM model in comparison to the $\approx1\%$ improvement with EW-MSE loss over MSE loss with GRU model. This is likely because of LSTM's complex architecture, which  allows it to capture long-term dependencies more effectively. While GRU offers faster training and competitive performance, LSTM can capture longer-term dependencies, with the choice often depending on compute constraints and task needs.

\subsection{Scalability Study}
In Figs.~\ref{fig:MSEvsEWMSE} and~\ref{fig:GRUvsLSTM} we also demonstrate the scalability of the FL-LSTM model for CA~(tested on unseen 39,291 buildings), FLO~(tested on unseen 12,344 buildings) and RI~(tested on unseen 1,276 buildings). It is important to note that the LSTM model trained in a federated fashion is directly deployed for testing on large unseen samples without any re-training at the client side, highlighting the generalization of the FL model. Deployment of such cloud models to clients that have minimal or no compute bandwidth available for training is very apt for practical applications. The FL-LSTM model trained on 120 buildings performs well even when deployed on a large, unseen test set.

\subsection{FL Training on Raspberry Pi Edge Cluster}
Finally, to underscore the low resource usage and practical deployability of our proposed FL method, we provide brief results of FL training using Flotilla~\cite{banerjee2025flotilla} on a Raspberry Pi cluster, comparable to edge  compute available in smart power meters. The FL server is hosted on a AMD Ryzen 9 3900X 12-Core CPU workstation and each FL client is on a Raspberry Pi 4B with a 4-core ARM A72 CPU at 1.8GHz, having 2--8GB RAM, connected over Gigabit Ethernet.
We randomly sample 30 buildings from the 100 buildings in the \S\ref{subsec:restults-clustering}. We train the LSTM model for 100 FL rounds, selecting all 30 clients in every round. This converges to an average of $93.2\%$ accuracy and $\approx10^{-3}$ loss. It takes a per-round training time of just  $70$--$100$~secs on the Raspberry Pis and model data transfer of 560KB over the network, with 450MB of memory footprint. This offers promise to practically deploy such lightweight FL training in the field using energy monitoring edge devices.

\section{Conclusion}
This paper presents a methodical study of federated learning for demand forecasting in micro-grids using edge and cloud resources. The clustering and loss functions, and the DNN and ARIMA models we study, offer a detailed view of the impact of these configuration parameters of the prediction accuracies. The validation done on three states and 1000s of held-out buildings while training on just 120 indicates the robustness of these privacy-preserving FL methods for heterogeneous data. The proposed method demonstrates the ability to lower compute and communication costs on edge and cloud while scaling to large cities and micro-grids. 

Overall, these findings offer valuable insights into the effective deployment of federated learning for demand prediction in diverse energy systems in smart cities and microgrids using edge and cloud resources.
As future work, we plan to validate these on real-world deployment in our campus microgrid, and also extend these to other campus utilities.

\bibliographystyle{plain}
\bibliography{arxiv-main}

\noindent\footnotesize\textbf{Disclaimer:} This content is provided for general informational purposes only and is not intended to serve as a substitute for consultation with our professional advisors. The intellectual contribution is the property of Accenture and its affiliates, and Accenture holds the copyright and all associated intellectual property rights. No part of this document may be reproduced in any form without prior written permission from Accenture. The opinions expressed herein are subject to change without notice.
\end{document}